\documentclass[prb,superscriptaddress,twocolumn,showpacs,floatfix]{revtex4-1}
\usepackage{amsmath}
\usepackage{amsfonts}
\usepackage{amssymb}
\usepackage{graphicx,graphics,color,%epsfig
}

\newcommand{\beq}{\begin{equation}}
\newcommand{\eeq}{\end{equation}}
\begin{document}

%\title%{The $T^{\ast }$-line in the phase diagram of the quantum critical heavy fermion compound YbRh$_{2}$Si$_{2}$ : The onset of spin-flip scattering%
\title{Spin-flip scattering of critical quasiparticles and the phase diagram
of YbRh$_2$Si$_2$
}
\author{Peter W\"{o}lfle}
\affiliation{Institute for Theory of
Condensed Matter and Institute for Nanotechnology, Karlsruhe Institute of Technology, 76049 Karlsruhe,
Germany}
\author{Elihu Abrahams}
\affiliation{Department of Physics and Astronomy, University of California
Los Angeles, Los Angeles, CA 90095} 

\date{\today}

\begin{abstract}
Several observed transport and thermodynamic properties of the heavy-fermion compound YbRh$_{2}$Si$_{2}$ in
the quantum critical regime are unusual and suggest that the fermionic quasiparticles are critical,
characterized by a scale-dependent diverging effective mass. A theory based on the concept of critical quasiparticles (CQP) scattering off
antiferromagnetic spin fluctuations in a strong-coupling regime has been shown to successfully explain the unusual existing data and
to predict a number of so far unobserved properties. In this paper, we
point out a new feature of a magnetic field-tuned quantum critical point of
a heavy-fermion metal: anomalies in the transport and thermodynamic
properties caused by the freezing out of spin-flip scattering of critical
quasiparticles and the scattering off collective spin excitations. We show
that a step-like behavior as a function of magnetic field of {\it e.g.} the Hall
coefficient and magnetoresistivity results, which accounts
quantitatively for the observed behavior of these quantities. That behavior has been described as a crossover line $T^*(H)$
 in the $T-H$ phase diagram of YbRh$_{2}$Si$_2$. Whereas some authors have interpreted this observation as signaling the
breakdown of Kondo screening and an associated abrupt change of the Fermi
surface, our results suggest that the $T^{\ast }$ line may be quantitatively
understood within the picture of robust critical quasiparticles.
\end{abstract}
\pacs{}
\maketitle

\preprint{}
\affiliation{Institute for Theory of
Condensed Matter, Karlsruhe Institute of Technology, 76049 Karlsruhe,
Germany}

\affiliation{Department of Physics and Astronomy, University of California
Los Angeles, Los Angeles, CA 90095} 

\section{Introduction}
Quantum phase transitions in heavy-fermion compounds have attracted
considerable interest over the last two decades. These systems exhibit
deviations from the standard Fermi-liquid description of metals, as a consequence of the
interaction of the fermionic (Landau) quasiparticles with bosonic critical
spin fluctuations. The existence of phase transitions in these systems was
proposed early on by Doniach \cite{Doniach}, who argued that the competition
of Kondo screening of the local moments and the (RKKY) interaction between
them should lead to a quantum phase transition separating a
paramagnetic from a (usually) antiferromagnetic phase. A
full explanation of just how this happens is still lacking (for a review see Ref.\ \onlinecite{LRVW}). Therefore the discovery of a well-accessible quantum-critical regime in some heavy-fermion compounds has generated a good deal of research activity. In particular, YbRh$_{2}$Si$_{2}$ (YRS), which has a magnetic field-tuned quantum phase transition, has been studied extensively.  Motivated by experimental observations of
deviations from conventional quantum-critical behavior\cite{LRVW} at very low temperature in YRS, we have previously considered the interplay of fermionic (quasiparticle) and bosonic (spin fluctuations) critical behaviors and shown how this leads to {\it critical quasiparticles} with unconventional behavior in the critical region.\cite{wa11} The behavior of several of the transport and thermodynamic properties in the critical regime were successfully accounted for on the basis of the critical quasiparticle theory, both in YRS\cite{wa11,aw12} and CeCu$_{6-x}$Au$_x$.\cite{asw14}   In addition to the unconventional behavior in YRS as  $T\to 0$,  a crossover behavior in the Hall
constant \cite{Silke} and several other quantities \cite{sci} was observed along a line $T^{\ast }(H)$ in the temperature ($T$) - magnetic field ($H$) phase diagram, with the crossover width scaling as $T$. This $T^{\ast }$ line begins at relatively high $(T,H)$,
monotonically decreases with decreasing magnetic field and apparently
ends at the quantum critical point (QCP), which is accessed by tuning $H$
to the critical field $H_{c}\approx 0.06$ T. A number of authors have
interpreted this crossover as a signature of the breakdown of Kondo
screening and a concomitant change of the Fermi surface \cite{Si,pc,SVS}.

%The initial interpretation of the observed step-behavior in the Hall constant as a direct measure of the Fermi volume change turned out to be untenable, as the Hall constant has contributions from both electron and hole bands, and  therefore depends sensitively on the scattering properties of quasiparticles in these bands.
%This complex situation shows e.g. in the fact that the "jump" of the Hall constant decreases with decreasing temperature to almost zero.

%However, an interpretation of the observed step-like behavior of the Hall constant as a direct measure of the Fermi surface volume is not without complexities, inasmuch as the Hall constant has contributions from both electron and hole bands \cite{bs,zw}, and therefore depends sensitively on the scattering properties of quasiparticles in these bands.

In this paper, we propose an alternative explanation for the $T^{\ast }$ line that is based on the theory \cite{wa11,aw12} of well-defined critical quasiparticles with robust Kondo screening. It was originally developed for disordered systems, in which impurity scattering serves to distribute to the entire Fermi surface the critical renormalization of the quasiparticle effective mass. This results from interaction with antiferromagnetic spin fluctuations that in the pure case is important only at ``hot spots" on the Fermi surface.  Later, it was shown that exchange of pairs of AFM spin
fluctuations, ({\it i.e.} energy fluctuations), carrying small total momentum
leads to critical quasiparticles even for clean systems.\cite{asw14} In both cases, there results
scaling behavior of the free energy and transport properties that is
characterized by fractional power laws in temperature and in the tuning
parameter. In renormalization group language, the theory has two stable fixed
points at weak and strong coupling. The predictions of this theory were found
to be in excellent agreement with all available experimental data. 

We
argue here that the thermal activation of spin-flip excitations of critical quasiparticles in a non-zero
magnetic field leads to a threshold behavior of transport properties as a
function of magnetic field at fixed temperature. There are actually two
types of processes contributing to this threshold behavior, which defines a crossover $T^*(H)$,  Near the QCP,
the switching on of spin-flip scattering leads to a step-like feature in the
imaginary part of the critical component of the quasiparticle self energy
(at a temperature $T_{1}^{\ast }(H))$, which by analyticity carries over to
the real part and hence to the thermodynamic properties. The properties of
this $T_{1}^{\ast }$ line mirror those of the experimentally determined $%
T^{\ast }(H)$ line near the QCP. As we show below, the $T_{1}^{\ast }$ line
approaches the QCP following a fractional power law. In the temperature
regime for which data are available at present, the asymptotic low temperature
behavior has not yet been reached. Nonetheless, an evaluation of the $T_{1}^{\ast
}$ line using the available thermodynamic data approaches the QCP almost
vertically in the $T-H$ phase diagram, as apparently observed for $T^*(H)$. This effect
arises as a consequence of the unusual renormalization of the bare single
quasiparticle Zeeman splitting $h_{0}=g\mu _{B}H$. As we explain below, the magnetic field is
screened by the Fermi liquid interaction, such that $h_{0}\rightarrow
h=R_{nl}h_{0}$, where $R_{nl}(T,H)$ is a generalized Wilson ratio (in the
limit $H\rightarrow 0$, $R_{nl}\rightarrow R$, where $R$ is the
usual Wilson ratio). This renormalization of the Zeeman splitting played an
important role in the interpretation of the linewidth of electron spin
resonance (ESR) in YRS \cite{esr1,esr2}, in that it increases the low
temperature linewidth in the Fermi liquid regime by two orders of magnitude,
in agreement with experiment.\cite{sichel} Since the Wilson ratio is found to tend to
zero upon approaching the QCP, the  renormalized Zeeman splitting of the quasiparticle
energy is predicted to nearly vanish (the observed finite response to a magnetic
field is accounted for by the noncritical, nonquasiparticle contribution) .
Consequently, the thermal energy required to flip the spin nearly vanishes as $%
H\rightarrow H_{c}$ and $T\rightarrow 0$, leading to a threshold anomaly at $%
T_{1}^{\ast }(H)$ the width of which goes to zero as well. At higher $T$ the
anomaly is rapidly washed out.

There exists, however, a second type of spin-flip excitation, the collective
excitation observed in ESR experiments \cite{sichel,schaufuss}. This excitation has
also been seen in inelastic neutron scattering experiments \cite{Broholm}.
The condition of thermal energy being equal to the ESR energy quantum $%
\omega _{r}$ defines a line $T_{2}^{\ast }(H)$ which we find coincides with the
experimentally determined $T^{\ast }$ line at higher $T,H$. At lower field
the $T_{2}^{\ast }$ line crosses the critical field at a non-zero temperature
and therefore with a non-zero width. We therefore find a crossover behavior
near the point where $T_{1}^{\ast }(H)$ and $T_{2}^{\ast }(H)$ meet.

In Sec.\ II, we review the reasons leading to a renormalization of
the single-particle Zeeman splitting. In Sec.\ III, we calculate the imaginary part of the
self energy of the critical quasiparticles and derive the threshold behavior at the line $T_{1}^{\ast }(H)$.
This allows an approximate calculation of  a step-like feature in the
magnetoresistivity in Sec.\ IV and in the Hall coefficient in Sec.\ V, which is compared with
experiment. Then we deduce the real part of the self energy using
analyticity arguments, and therefore the quasiparticle weight factor $Z$.
The effective mass ratio obtained from the relation $m^{\ast }/m=1/Z$ allows
the identification, in Sec.\ VI, of a step-like feature in the magnetic-field dependence
of the specific heat and other thermodynamic quantities.

In Sec.\ VII, we calculate the contribution of thermal excitation of the ESR
spin resonance to the imaginary part of the self energy. Using analogous
arguments as in Sec.\ II, we derive the threshold contributions to the
transport quantities along a line $T_{2}^{\ast }(H)$, defined by the
scattering off the spin resonance.

We collect the results in Sec.\ VIII and compare the theoretically determined 
$T^{\ast }(H)$ line with the published experimentally-determined $T^{\ast
}(H)$ and find excellent agreement.

We summarize our findings in Sec.\ IX and give a critical evaluation of the
interpretation of the $T^{\ast }$ line as corroborating the picture of
critical quasiparticles%

\section{Renormalization of the Zeeman splitting}

\subsection{ Fermi liquid theory}

The single particle Green's function in a magnetic field has the form

\begin{equation}
G_{\sigma }(k,\omega ;H)=\frac{1}{\omega -\epsilon _{k}+\sigma h/2-\Sigma
_{\sigma }(k,\omega ;H)}
\end{equation}%
Here $h=R_{nl}h_{0}$ and $h_{0}=g\mu _{B}H$ where $R_{nl}$ is the renormalized Wilson ratio formulated as follows: The external field is screened by the molecular field 
$h_{0}\rightarrow h_{0}(1-f^{a}M/H),$ where $M=\sum_{\sigma }\int d\omega\int
d^{d}k(2\pi )^{-d}\sigma G_{\sigma }(k,\omega ;H)$ is the spin polarization
and $f^{a}$ is the (Landau quasiparticle) spin exchange interaction. In this paper, we consider a three-dimensional metal, with critical fluctuations also in $d=3$. In the limit $H\rightarrow 0$
, or more generally, if $M$\ is linear in $H$, we have $M=\chi H$. Using
the Fermi liquid expression for the spin susceptibility (in appropriate
units) $\chi =\partial M/\partial H=N_{0}^{\ast }/(1+F^{a})$, where 
$F^{a}=N_{0}^{\ast }f^{a}$ is the Landau parameter in the spin channel and $%
N_{0}^{\ast }$ is the quasiparticle density of states, we then get $%
h=h_{0}/(1+F^{a})=Rh_{0}$ , where $R=\chi /N_{0}^{\ast }=1/(1+F^{a})$ is the
usual Wilson ratio. Now, in the case of non-zero magnetic field $R_{nl}=1-f^{a}\chi
b$, where $b=(M/H)/\chi $. Here $\chi =\partial M/\partial H$ is the
differential susceptibility at finite field $H.$ Expressing $f^{a}\chi
=F^{a}/(1+F^{a})=1-R$, we finally get

\begin{equation}
h=h_{0}[1+(R-1)b]=h_{0}R_{nl}  \label{hmolfield}
\end{equation}%
The static screening changes the applied field $h_{0}$ to $h$
everywhere, so that we shall use the screened field in place of the bare
field from now on. The screening factor $R_{nl}$ ($nl$ stands for ``nonlinear
screening") is expressed in terms of the Wilson ratio $R$ and $b$, the ratio of nonlinear and differential susceptibility. Here, $R =\alpha
_{R}\chi T/C$, with $C/T=\gamma$, the specific heat coefficient, $\alpha_R=(2\pi k_B/3g\mu_B)^2$,  and $g=3.6$ is the
g-factor. The experimental data show that $b>1$, always. and $b(H)$ is an increasing
function of $H$, since the slope of $%
M(H)$ becomes smaller for increasing $H-H_{c}$. The nonlinearity of $M$ thus weakens the increase of $R$
with decreasing field towards $H_{c}$ .

Expanding the dynamic part of the self energy at small $\omega $ and
defining the quasiparticle (qp) $Z$-factor as $Z^{-1}=[1-\partial \Sigma
(k,\omega ;0)/\partial \omega ]$ we find 
\begin{equation}
G_{\sigma }(k,\omega ;h)=\frac{Z}{\omega -\epsilon _{k}^{\ast }+Z_{h\sigma
}(\sigma h/2) + i\Gamma},
\end{equation}%
where $\Gamma = Z{\rm Im}\Sigma$ and
\begin{eqnarray}
Z_{h\sigma } &=&\frac{1-\sigma \lbrack \Sigma _{\uparrow }(k,0;h)-\Sigma
_{\downarrow }(k,0;h)]/h}{[1-\partial \Sigma (k,\omega ;0)/\partial \omega
]_{\omega =0}} \\
\epsilon _{k}^{\ast } &=&Z[\epsilon _{k}+\frac{1}{2}(\Sigma _{\uparrow
}(k,0;h)+\Sigma _{\downarrow }(k,0;h))]
\end{eqnarray}%
In the limit of $h\rightarrow 0$\ we may use the relation 
\begin{equation}
\lim_{h\rightarrow 0}2\sigma \partial \Sigma _{\sigma }(k,0;h)/\partial
h=\lim_{\omega \rightarrow 0}\partial \Sigma (k,\omega ;0)/\partial \omega,
\label{h-ident}
\end{equation}%
to find $Z_{h\sigma }=1$. So, the coupling of qp spins to the external
field is only renormalized by the molecular field. This is in accord
with the statement that Landau quasiparticles have the same quantum numbers
as bare particles, and therefore the qp spin is a conserved quantity.

\subsection{Renormalization of the Zeeman splitting of critical
quasiparticles near a field-tuned QCP}

The above relation of the two derivatives of the self energy with respect to 
$h$ and $\omega $ does not hold generally in non-zero magnetic field. This can
be seen by analyzing any diagram of the self energy $\Sigma _{\sigma
}(\omega )$ in terms of bare Green's functions in the following way: there
is always exactly one string of Green's functions $G_{\sigma }(k_{j},\omega
-\omega _{1}-...)=(\omega +\sigma h/2-\omega _{1}-...-\epsilon _{k_{j}})^{-1}
$ (carrying the external spin label) connecting beginning and end of the
diagram. In those Green's functions a shift of magnetic energy $\sigma h$ is
equivalent to a shift of $\omega $ . All other Green's functions belong to
closed loops in which the spin index is summed over. The closed loop
contributions are then necessarily functions of $H^{2}$. In the limit $%
H\rightarrow 0$ those $H^{2}$ corrections drop out. Hence in this limit the
relation Eq.\ (\ref{h-ident}) holds. At non-zero field the $H^{2}$ corrections
are not negligible (although they may be small of $O(H/\epsilon _{F})^{2}$,
where $\epsilon _{F}$\ is the Fermi energy) and Eq.\ (\ref{h-ident}) does
not hold in general. However, near the critical field $h_{c}$ the derivative 
$\partial \Sigma _{\sigma }(k,0;h)/\partial h$ is critically enhanced. As
suggested in Ref. \onlinecite{asw14}, the vertex function corresponding to the
derivative $\partial \Sigma _{\sigma }(k,0;h)/\partial h$ is enhanced $%
\propto 1/Z$\ ,  diverging at the QCP just like $\partial \Sigma (k,\omega
;0)/\partial \omega $.\ Therefore, the relation Eq.(\ref{h-ident}) still
holds, as far as the critical contributions are concerned, and,
as explained in the previous subsection, we have

\begin{equation}
h=R_{nl}h_{0}+h_{reg}.
\end{equation}%
In the following we shall drop the regular contribution $h_{reg}$, since it
vanishes at the critical point faster than the first contribution, at least $%
\propto Z^{2}$ .

The relation of the renormalized Zeeman splitting $h$ to the magnetic field $%
H$ is somewhat involved and in general may not be expressed as a simple
functional relationship. Near the QCP of YRS, however, we may use the result \cite{aw12}
$Z^{-1}\propto (H-H_{c})^{1/3}\propto h$, from which follows $T^{\ast
}(H)\propto $ $(H-H_{c})^{2/3}$ as the asymptotic form of the $T^{\ast }$ line.
At the temperatures for which data are available, the behavior of the magnetic susceptibility,
in particular, as a function of $T,H$ is not well-represented by a simple scaling form. It is then more reliable to use
directly the available experimental information on the specific heat
coefficient $\gamma (T,H)$ (see Ref. \onlinecite{oesch}), the differential spin
susceptibility $\chi (T,H)$ (see Refs. \onlinecite{trov,geg02,geg06,cust}), and the
magnetization $M(T,H)$ (see Refs. \onlinecite{tok09,geg06,bran}) to determine the
quasiparticle Zeeman splitting $h(T,H)$. In Fig.\ 1 we show results for
 $h(T,H)$ versus magnetic field $H$ at four selected
temperatures $T=18,38,65,100$ mK, for which magnetoresistivity data,\cite{fried10} Sec.\ IV below, are
available.  The $T=18$K data is shown as the purple line while the  other three temperatures give similar results, as shown. Also shown is the bare Zeeman splitting $h_{0}(H)$. One can see
that $h$ is substantially enhanced by the ferromagnetic molecular field. At
these temperatures the asymptotic behavior $h\propto (H-H_{c})^{1/3}$
mentioned above is not yet seen. It is masked by the relatively strong
increase of the spin susceptibility towards lower fields (at these low
temperatures critical antiferromagnetic spin fluctuations dominate and $\chi 
$ eventually reaches its asymptotic $T=0$ limiting value).

%\begin{figure}
\begin{figure}[h]
\centering
\includegraphics[width=.51\textwidth, %viewport=0 40 600 400,%clip
]
{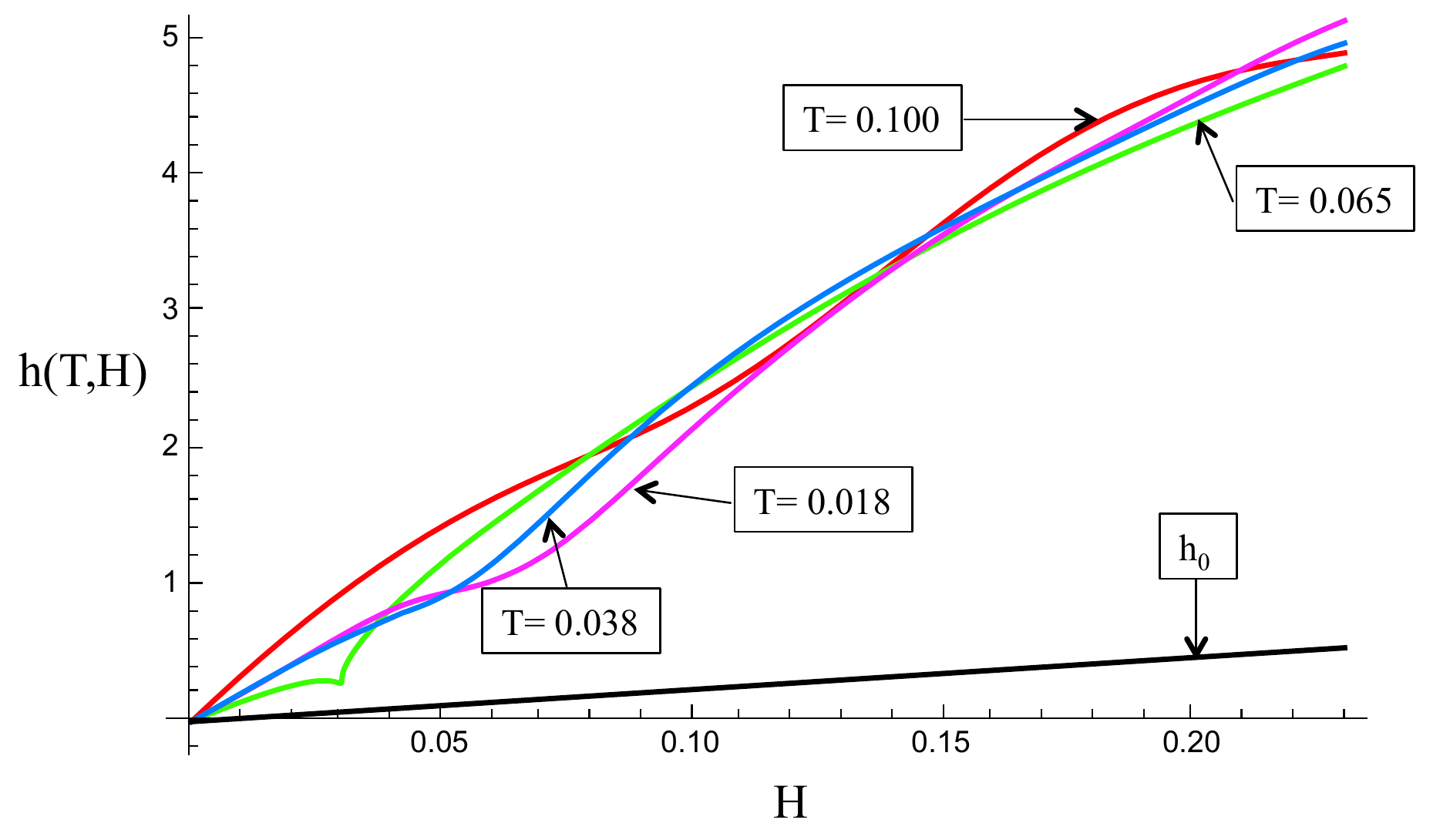}
%\vskip -1.5cm
\caption{Renormalized Zeeman splitting $h(T,H)$ at various $T$, in degrees K at various $T$. %$T=18$ mK is shown explicitly, $T=$ 38, 65 and 100 mK fall in the gray strip. 
The unrenormalized Zeeman splitting $h_0$ is also shown, The magnetic field unit is Tesla and the temperatures are in Kelvin.}\label{renormZ}
\end{figure}

\section{Threshold behavior of spin flip scattering of single
quasiparticles: self energy}

Since the differential spin susceptibility $\chi $ and the magnetization $M$
both approach a constant finite value at the QCP, and since the specific
heat coefficient $C/T$ as well as $Z^{-1}$\ diverge as $T^{-1/4}$ (in 3d - for 2d
AFM fluctuations, see  Ref.\ \onlinecite{asw14}), $h$ is seen to nearly vanish at the QCP (more precisely, $h/h_{0}=1-b$ at the QCP, which is a very small quantity). In
one-loop approximation of the self energy the most important effect of
Zeeman splitting is from the intermediate quasiparticle line (straddled by
a fluctuation propagator). The effect on the AFM spin fluctuation
propagator is small ($h$ does not enter the Landau damping term in lowest
order).

Let us then look at the self energy expression (due to coupling to magnetic energy fluctuations, as in 
Ref.\ \onlinecite{asw14}):

\begin{eqnarray}
{\rm Im}\Sigma_{\sigma}(k,\omega ) &\approx &\lambda _{E}^{2}\sum_{\sigma
^{\prime }}\int (d\mathbf{q})\int_{-\infty }^{\infty }d\nu F(\nu ,\omega )  \notag \\
&&\times {\rm Im}\chi ^{E}_{\sigma\sigma'}(\mathbf{q},\nu ){\rm Im}G_{\sigma'}(%
\mathbf{k+q},\nu +\omega ) \label{self}
\end{eqnarray}%
where $F(\nu ,\omega )=f(\nu +\omega )+b(\nu )$\ and $f(\omega
)=1/(e^{\omega /T}+1)$ and $b(\omega )=1/(e^{\omega /T}-1)$. The energy
fluctuation spectrum is given by (see Ref.\ \onlinecite{asw14}):

\begin{equation}
{\rm Im}\chi ^{E}(\mathbf{q},\omega )\propto \frac{\omega (|\omega
|/Z^{2})^{3/2}}{(q^{2}+\xi ^{-2}+|\omega |/Z^{2})^{2}}
\end{equation}%
The wavevector $\mathbf{q}$ and the inverse correlation length $\xi ^{-1}$  are in
units of the Fermi wavenumber $k_{F}$ and the fluctuation energy $\omega $
as well as all other energies ($T,h$)\ in units of the Fermi energy $%
\epsilon _{F}$, of the heavy quasiparticle band of
YRS, approximately 10 K. The decisive effect of the Zeeman splitting is
on the result of the angular integration over $\mathbf{q}$ in Eq.\ (8).
\begin{align}
&\int \frac{d\Omega _{q}}{4\pi }{\rm Im}G_{\sigma ^{\prime }}(\mathbf{k+q}%
,\omega ) \notag \\
\approx &\int \frac{d\Omega _{q}}{4\pi }Z\delta (\omega -\epsilon _{k\sigma
^{\prime }}^{\ast }-v_{F}^{\ast }q\cos \theta )  \notag \\
 =&\frac{Z}{v_{F}^{\ast }q}\theta (v_{F}^{\ast }q-|\omega -\epsilon
_{k\sigma ^{\prime }}^{\ast }|)  \label{ImG}
\end{align}%
where $v_{F}^{\ast }=Zv_{F}$\ and $v_{F}$\ are the quasiparticle and bare
Fermi velocities. We need ${\rm Im}\Sigma _{\sigma }(k,0)$
at the Fermi energy, {\it i.e.} $\epsilon _{k\sigma }^{\ast }=0$ and $%
\epsilon _{k\sigma ^{\prime }}^{\ast }=h(\sigma ^{\prime }-\sigma )/2$.
Also, $\omega \ll $ $v_{F}^{\ast }q$, as may be seen from the structure of $%
{\rm Im}\,\chi ^{E}(\mathbf{q},\omega )$, so that $\omega$ may be dropped
in Eq.\ (\ref{ImG}).\ We now see that the non-spinflip term $\sigma ^{\prime
}=\sigma $ gives rise to half of the contribution we had previously (at $H=0$). The spin-flip term, however, has the additional constraint on the $q$%
-integration $v_{F}^{\ast }q>|h|$. The $q$-integral of the spin flip term in Eq.\ ({\ref{self})
may be approximated by $\Phi (\omega ;\xi ,T,h)$ defined as

%\begin{align}
%&\int (dq){\rm Im}\chi _{E}(\mathbf{q},\omega )%
%\frac{1}{v_{F}q}\theta (v_{F}^{\ast }q-h)  \notag \\
%& \approx Z^{-3}\int \frac{dq^{2}\omega ^{5/2}}{(q^{2}+\xi ^{-2}+|\omega
%|/Z^{2})^{2}}\theta (q^{2}-h^{2}/Z^{2}v_{F}^{2})  \notag \\
%& \approx \frac{\omega |\omega |^{3/2}}{Z^{3}}[\frac{\theta (Z^{2}\xi
%^{-2}+|\omega |-h^{2})}{\xi ^{-2}+|\omega |/Z^{2}}+\frac{\theta
%(h^{2}-Z^{2}\xi ^{-2}-|\omega |)}{h^{2}/Z^{2}}]
%\end{align}%

\begin{align}
&\Phi (\omega ;\xi ,T,h)= \int q^2dq\,{\rm Im}\chi _{E}(q,\omega )%
\frac{1}{v_{F}q}\theta (v_{F}^{\ast }q-h)  \notag \\
& \approx Z^{-3}\int \frac{(qdq)\omega ^{5/2}}{(q^{2}+|\omega
|/Z^{2})^{2}}\theta (q^{2}-h^{2}/Z^{2}v_{F}^{2})  \notag \\
& \approx \frac{\omega |\omega |^{3/2}}{Z^{3}}\left[\frac{\theta 
(|\omega |-h^{2})}{|\omega |/Z^{2}}+\frac{\theta
(h^{2}-Z^{2}-|\omega |)}{h^{2}/Z^{2}}\right]
\end{align}
where $\theta (x)$ is the unit step function and we have taken $\xi \to 0$, since we restrict ourselves to the critical region
%In the critical regime, where we may take $\xi ^{-1}\rightarrow 0$, we get,
We use this result in Eq.\ (\ref{self}) to obtain

\begin{align}
{\rm Im}\Sigma _{\sigma }^{cr}(\omega )&\approx \lambda
_{E}^{2}\int_{-\infty }^{\infty }d\nu F(\nu ,\omega )[\Phi (\nu ;T,0)+\Phi
(\nu ;T,h)] \notag \\
& \approx aT^{3/4}[K(0,\omega /T)+K(u/T,\omega /T)],  %\notag
\label{ims}
\end{align}%
where $u=h^2/\epsilon_F $ and we defined a function $K(z,y)$ describing the scaling behavior as the
magnetic field\ and the energy $\omega $ is varied as follows:
\begin{eqnarray}
K(z,y) &=&I(z,y)/I(0,0)\label{kzy} \\
I(z,y) &=&I_{3/2}(z,y)+[I_{5/2}(0,y)-I_{5/2}(z,y)]/z \\
I_{\mu }(z,y) &=&\int_{z}^{\infty }dxF(x,y)x^{\mu }\label{ii}
\end{eqnarray}%
where the thermal factor $F(x,y)$ is defined below Eq.\ (\ref{self}). We also used $\lambda
_{E}\propto Z^{-3}$ and we approximated $Z$ by $Z(\omega =0;\xi
\to\infty ,T,H)= {\rm const.}(T/T_{0})^{1/4}$ .

The function $K(h^{2}/\epsilon _{F}T,0)$ drops monotonically with increasing 
$h$, in a step-like fashion. The step width scales with temperature $T$.\
This means that the spin flip term shows threshold behavior as a function of 
$h$ at $h\approx (\epsilon _{F}T)^{1/2}$ . As mentioned above, the
dependence of $h$ on magnetic field does not follow a simple functional form,
so that the $T^*$ line has to be determined numerically.

In the following we will use the above result for the self energy to
determine the contribution of qp spin-flip scattering to several of the
quantities for which a threshold behavior as a function of $H$ has been
observed. In particular we will give a detailed comparison of the
magnetoresistance and Hall coefficient data with our theoretical result.

\section{Magnetoresistivity}

%The magnetoresistivity, at low  enough temperature that impurity scattering of rate $1/\tau ^{imp}$ dominates the imaginary part of the self energy, is obtained by averaging the quasiparticle transport relaxation rate $1/\tau _{tr}^{\ast }(\omega)=Z(\omega ){\rm Im}\Sigma _{\sigma }(k,\omega )$ over energy  from the Kubo formula for the magnetoconductivity by using the appoximation $1/\tau^* = Z(\omega){\rm Im}\Sigma \ll 1/\tau^{imp}$
%(here $f$ is the Fermi function)

The magnetoresistivity $\rho(T,H)$ is determined by the quasiparticle scattering rate due to impurities $1/\tau^{imp}$ and that due to scattering off the critical fluctuations $1/\tau^* = Z(\omega){\rm Im}\Sigma(\omega)$. At low enough temperature, impurity scattering dominates. We obtain  $\rho(T,H)$ from the Kubo formula for the conductivity by expanding in the small quantity $\tau^{imp}/\tau^* \ll 1$ as follows ($f$ is the Fermi function):
\begin{eqnarray}
\rho (T,H)-\rho (0,H)) &=&\left[ \sum_{\sigma }\int d\omega \frac{\partial f%
}{\partial \omega }\frac{N_{0}^{\ast }(\omega )v_{F}^{\ast 2}(\omega )(\tau
^{imp})^{2}}{\tau^{\ast }(\omega )}\right]   \notag \\
&&\times \left[ \sum_{\sigma }\int d\omega \frac{\partial f}{\partial \omega 
}N_{0}^{\ast }(\omega )v_{F}^{\ast 2}(\omega )\tau _{imp}\right] ^{-2} 
\notag \\
&=&a\sum_{\sigma }\int d\omega \frac{\partial f}{\partial \omega }{\rm Im}%
\Sigma _{\sigma }(k,\omega )  \label{magres_gen}
\end{eqnarray}%
where the renormalized density of states $N_{0}^{\ast }(\omega )=N_{0}/Z$,
the Fermi velocity $v_{F}^{\ast }=v_{F}Z$, and the qp relaxation rate $1/\tau^
{\ast }=  Z{\rm Im}\Sigma $\ , so that $N^{\ast
}(\omega )v_{F}^{\ast 2}(\omega )\propto Z(\omega )$, cancelling the
factor of $Z$ in $1/\tau^*$, the quasiparticle relaxation rate (and we neglect a
contribution from vertex corrections, which may be assumed to change only
the prefactor).\cite{wa11,aw12,asw14} Using Eq.\ (\ref{ims}) for ${\rm Im}\Sigma$
and scaling out the overall $T$-dependence, we find

\begin{eqnarray}
%\rho (T,H)-\rho (0,H) &\approx &a(m/e^{2}N_{0})(T/T_{0})^{3/4}L(u/T) 
\rho (T,H)-\rho (0,H) &\approx &a(m/e^{2}N_{0})T^{3/4}L(u/T) 
\label{magres} \\
L(z) &=&\int_{0}^{\infty }dy\frac{K(0,y)+K(z,y)}{\cosh ^{2}(y/2)}.
\label{Ldef}
\end{eqnarray}

In Sec.\ IIB, we found the renormalized Zeeman splitting $h$ numerically from the available experimental data (see Fig.\ 1). It enters the functions $K(z,y)$, Eq.\ ({\ref{kzy}) which determine the magnetoresistivity. Substituting the found renormalized Zeeman splitting into the expression for the
magnetoresistivity, we have evaluated Eq.(\ref{magres}) for four temperatures %
T=18,38,65, and 100 mK for which data are available. In Fig.\ 2, we compare our
results with the data for sample \#1 of Friedemann et al.\cite{fried10}
Here we approximated the magnetic field dependence of the background
resistivity by $\rho (0,H))=c_{1}+c_{2}H^{2}$\ . The characteristic energy $%
u=h^{2}/\epsilon_F$, where $\epsilon _{F}\approx 10$K and $h$ is obtained
from Eq.\ (\ref{hmolfield}), as shown in Fig.\ 1. The remaining unknown
parameter set \{$a \approx 1\mu \Omega {\rm cm},  c_{1}\approx 0.9\mu \Omega {\rm cm}, c_{2}\approx 0.5\mu \Omega {\rm cm}/{\rm T}^{2}$\} was chosen to give the
best fit to the data for all the temperatures chosen. As one can see, the
experimental data are described quite well by the theory, with the single set of parameters The width of the
step $\Delta H$\ is found to approximately scale with $T$ .

\begin{figure}[h]
\centering
\includegraphics[width=.5\textwidth, %viewport=30 80 600 10,%clip
]
{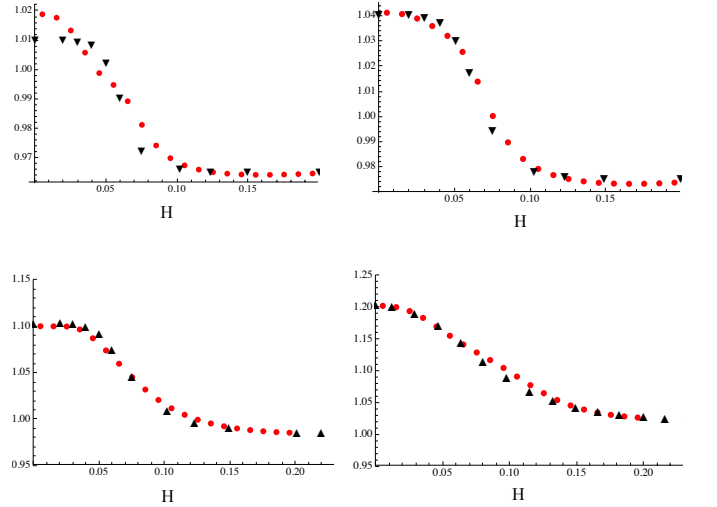}
%\vskip -1.5cm
\caption{Magnetoresistivity $\rho(T,H)$ at various $T$. Clockwise from upper left, $T=18,38,65$ and 100 mK. $H$ in Tesla and $\rho$ in $\mu\Omega$-cm. Dots are theory, Eq.\ (\ref{magres}); triangles are from the data, Ref.\ \onlinecite{fried10}}
\label{rMR}
\end{figure}

The agreement of our simplified model calculation with experiment is
remarkable, considering that we have adopted a number of approximations,
including the neglect of the $H$-dependence of $Z$ (assuming that the magnetic
field region considered here lies completely inside the critical regime, and
 neglect of the difference of effective masses of spin ($\uparrow ,\downarrow 
$)-quasiparticles.

\section{Hall coefficient $R_{H}$}

Electronic structure calculations within the ``renormalized band theory,"
{\it i.e.}  taking the Kondo resonance scattering at the Yb ions into account,
have revealed two relevant bands involved in transport, one of particle, the
other of hole character.\cite{zwick10} As a consequence,
substantial compensation is observed in the Hall coefficient data, leading
to small values of $R_{H}$ and\ an enhanced sensitivity to disorder.\cite%
{pasch04,fried10,zwick10} The Hall coefficient is given in terms of the
partial Hall ($\sigma _{xyz}^j$) and longitudinal ($\sigma _{xx}^j$)\
conductivities of the two bands ($j=1,2$) as \cite{zwick10}

\begin{equation}
R_{H}(T,H)=\frac{\sum_{j=1,2}\sigma _{xyz}^j}{(\sum_{j=1,2}\sigma
_{xx}^j)^{2}}  \label{RHall}
\end{equation}%
where

\begin{eqnarray}
\sigma _{xyz}^j &=&\sum_{\mathbf{k}}\tau _{\mathbf{k},j}^{\ast 2}u_{\mathbf{k%
},xy,j}^{\ast }\left( \frac{\partial f}{\partial \epsilon_{\mathbf{k}j}^{\ast }}%
\right)  \\
\sigma _{xx}^j &=&\sum_{\mathbf{k}}\tau _{\mathbf{k},j}^{\ast }v_{\mathbf{k}%
,x,j}^{\ast 2}\left( -\frac{\partial f}{\partial \epsilon_{\mathbf{k}j}^{\ast }}%
\right) 
\end{eqnarray}%
Here $u_{\mathbf{k},xy,j}^{\ast }=[v_{\mathbf{k},x,j}^{\ast }v_{\mathbf{k}%
,x,j}^{\ast }M_{yx,j}^{\ast -1}-v_{\mathbf{k},x,j}^{\ast 2}M_{yy,j}^{\ast
-1}]$, and $v_{\mathbf{k},x,j}^{\ast }$ and $M_{yx,j}^{\ast -1}$ are the $x$%
-component of the quasiparticle velocity and the $yx$-component of the
inverse quasiparticle mass tensor of the $j$-th band (as earlier, the asterix indicates the quasiparticle renormalization). As in the case of the
magnetoconductivity, we use the fact that the inelastic scattering from critical fluctuations gives
only a small contribution to the scattering rate, %$1/\tau _{\mathbf{k},j}^{\ast }=1/\tau _{\mathbf{k}j}^{\ast imp}+1/\tau _{\mathbf{k}j}^{\ast inel}$, 
thus allowing expansion in the small parameter $\tau _{\mathbf{k}j}^{imp}/\tau _{\mathbf{k}%
j}^{*}$: 
\begin{align}
\Delta \sigma _{xyz}^j(T,H) &=2\sum_{\mathbf{k}}\frac{(\tau _{\mathbf{k}%
,j}^{\ast imp})^{3}}{\tau _{\mathbf{k}j}^{\ast inel}}u_{\mathbf{k}%
,xy,j}^{\ast }\left( \frac{\partial f}{\partial \epsilon_{\mathbf{k}j}^{\ast }}%
\right)   \notag \\
&\propto (N_{0}v_{F}^{2}/m)(\tau_j ^{imp})^{3}u_{j} \notag \\
&\times \int d\omega \left( -\frac{\partial f}{\partial \omega }\right) {\rm Im}%
\Sigma _{j}(\omega ) 
\label{Hall-cond}
\end{align}%
where $\Delta \sigma _{xyz}^j(T,H)=\sigma _{xyz}^j(T,H)-\sigma _{xyz}^j(0,H)$. Here, we have used
 $u_{\mathbf{k},xy,j}^{\ast }(\partial
f/\partial \epsilon_{\mathbf{k}j}^{\ast })\rightarrow
u_{j}(v_{F}^{2}/m)Z_{j}^{2}(\omega )(\partial f/\partial \omega )$, and have
accounted for band structure effects in an average way by the dimensionless factor $%
u_{j}\lessgtr 0$. We recall that the impurity relaxation
rate in the case of unitary scattering (which we assume to be dominant) is
renormalized as $1/\tau _{\mathbf{k}_{F},j}^{\ast imp}$ $\propto Z(\omega
)/\tau _{\mathbf{k}_{F},j}^{imp}$. In the critical regime, using the
results obtained for $Z(\omega)$ and ${\rm Im}\Sigma (\omega )$ for YRS in
the regime dominated by three-dimensional antiferromagnetic fluctuations \cite%
{asw14}\ we may scale out the temperature dependence by using $Z_{j}(\omega
)\propto |\omega |^{1/4}$ and ${\rm Im}\Sigma _{j}(\omega )\approx \gamma
_{j}|\omega |^{3/4}$, $\gamma _{j}> 0$. As mentioned above, both data and theory suggest that the Hall coefficient $R_{H}(0,H)$\ at $T=0$, and
therefore $\sigma _{xyz}(0,H)=\sum_{j}\sigma _{xyz}^j(0,H)$, are rather
small, as particle and hole contributions almost compensate. The temperature-
dependent contribution may be approximated as
%\begin{align}
%R_{H}(T,H) &=[\rho (T,H)]^{2}[\sigma _{xyz}(0,H)+\sum_j \Delta \sigma_{xyz}^j(T,H)]  \label{RH} \\
%\Delta \sigma _{xyz}^j(T,H) &=a'_{H}\frac{n(\tau ^{imp})^{3}u}{%
%m^{2}}T^{3/4}L(w/T)
%\end{align}%
\begin{equation}
R_{H}(T,H) =[\rho (T,H)]^{2}[\sigma _{xyz}(0,H)+\sigma'
_{xyz}(T,H)],  \label{R'H}
\end{equation}
where, using Eq.\ (\ref{Hall-cond}),
\begin{align}
\sigma'(T,H) &= \sum_{j}\Delta\sigma _{xyz}^j(0,H)\notag \\ 
&= a'_{H}\frac{n(\tau ^{imp})^{3}b}{%
m^{2}}T^{3/4}L(u/T).
\end{align}

$L(z)$ was defined in Eq.\ (\ref{Ldef}).\ Here the dimensionless
quantity $b\propto (b_{1}\gamma _{1}+b_{2}\gamma _{2})$ , with $b_{1}>0$
(particles) and $b_{2}<0$ (holes) describes the extent of compensation. We
note that in the extreme limit of low temperature, when the inelastic
component of \ $\sigma _{xx}^j$ may be neglected, such that the denominator
of Eq.\ (\ref{RHall}) may be replaced by $[\rho (0,H)]^{-2}\propto (\tau
^{imp})^{2}$, two powers of $\tau ^{imp}$ in Eq.(\ref{Hall-cond}) are
cancelled and the $T$-dependent contribution to $R_{H}$ scales with disorder
strength as $\Delta R_{H}=R_{H}(T,H)-R_{H}(0,H)\propto \tau ^{imp}$ .

For the numerical evaluation of Eq.(\ref{R'H}) we used again a
parameterization of the impurity scattering contribution of the form $\sigma
_{xyz}(0,H)=c_{H1}+c_{H2}H^{2}$\ and defined $a_{H}=a_{H}^{\prime }n(\tau
^{imp})^{3}u/m^{2}$. In Fig.\ 3, we show a comparison of the calculated $R_{H}$ curves with experimental data,\cite{fried10}  again choosing a single set of parameters \{$c_{H1}\approx 1.7$, $c_{H2}=1.5$, $a_{H}\approx 0.5$\} and the
magnetoresistivity as determined above in Sec.\ IV. We conclude that the theory accounts
well for the observed behavior.

\begin{figure}[h]
\centering
\includegraphics[width=.5\textwidth, %viewport=30 80 600 10,%clip
]
{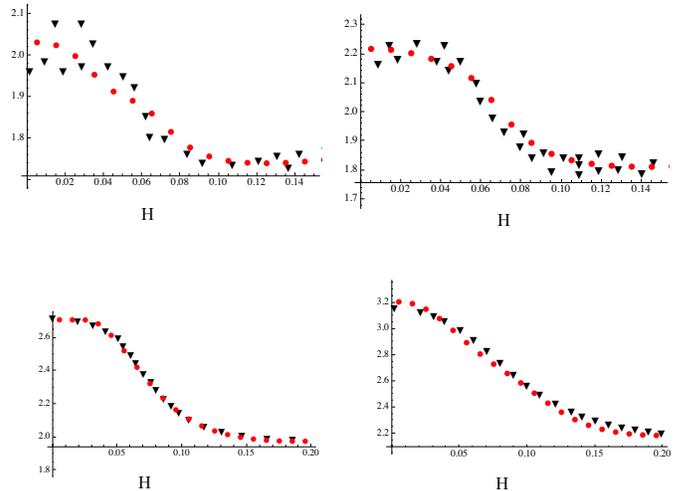}
%\vskip -1.5cm
\caption{Hall constant $R_H$ at various $T$. Clockwise from upper left, $T=18,38,65$ and 100 mK. $H$ in Tesla and $R_H$ in $10^{-10}{\rm m}^3/{\rm C}$. Dots are theory, Eq.\ (\ref{Hall-cond}); triangles are from the data, Ref.\ \onlinecite{fried10}.}
\label{hall}
\end{figure}

\section{Specific heat}

The specific heat coefficient is also affected, even though a bit weaker.
Using the analyticity properties of $\Sigma$, we have approximately

\begin{equation}
{\rm Re}\Sigma _{\sigma }(\omega )\approx (|\omega
|/Z^{2})^{7/2}1/|\omega| + 1/[|\omega|+h^2/\epsilon_F]
\end{equation}%
Since $Z^{-1}(T) = 1-\partial\Sigma/\partial\omega|_{\omega\to T}$, we get the self-consistent equation for $Z(T)$ as

\begin{equation}
Z^{-1}=1+Z^{-7}|T|^{3/2}[1+T/(T+h^{2}/\epsilon _{F})]
\end{equation}%
The strong coupling solution of this equation is

\begin{equation}
Z(T,H)\propto |T|^{1/4}[1+T/(T+h^{2}/\epsilon _{F})]^{1/6}
\end{equation}%
such that the specific heat coefficient would be

\begin{equation}
\gamma \propto T^{-1/4}[1+T/(T+h^{2}/\epsilon _{F})]^{-1/6}
\end{equation}%
This expression for the specific heat coeffient also describes step-like
behavior as the magnetic field is lowered through the $T^{\ast }$-line. We
do not attempt a detailed comparison with experiment here because the data
situation is not as good as in the case of the magnetoresistance.

\section{Scattering of quasiparticles by spin resonance bosons}

ESR experiments on YRS have shown a well-defined spin resonance \cite{sichel,schaufuss} in a wide region of the phase diagram extending from
fields as high as 8T down to the critical field (see, for example, Fig.\ 1
in Ref.\ \onlinecite{esr2}). Inelastic neutron scattering experiments \cite{Broholm}
have shown the existence of the ESR resonance at small but non-zero
momentum. Extending our earlier results \cite{esr1,esr2} to non-zero $q$, we
are led to the spin-fluctuation spectrum
\begin{align}
\mathrm{Im}\chi (q,\omega )& ={\rm Im}\left[\chi _{0}\frac{-\omega
_{r}-aq^{2}+i\gamma }{\omega -\omega _{r}-aq^{2}+i\gamma }\right]  \notag \\
& =\chi _{0}\frac{\omega \gamma }{(\omega -\omega _{r}-aq^{2})^{2}+\gamma
^{2}}  \notag \\
& \approx \chi _{0}\,\omega \delta (\omega -\omega _{r}-aq^{2})
\end{align}%
where $\omega _{r}$ is the spin-resonance frequency as calculated by us, see
Eq.\ (2) of Ref.\ \onlinecite{esr2} and $\gamma $ is the line width, see Eq.\ (4) and
following, of Ref.\ \onlinecite{esr2}

The resonance frequency $\omega _{r}$ is everywhere non-zero; it does not vanish
at the QCP, but gets renormalized to about $2/3$ its high field limiting
value in the critical regime.\cite{esr2} The coefficient $a$ has some
$T$-dependence that we neglect. Since the resonance linewidth is found
to be much less than the resonance frequency, we may take it to be
infinitesimally small. The imaginary part of the electron self energy caused
by scattering on the spin resonance is then given by
\begin{align}
{\rm Im}\Sigma (\omega ) &=\lambda ^{2}\int_{-\infty }^{\infty }d\nu
F(\nu ,\omega )\chi _{0}\nu \int d{\bf q}  \notag \\
&\times \mathrm{Im}G(\nu +\omega ,%
\mathbf{k+q)}\delta (\nu -\omega _{r}-aq^{2}),
\end{align}
Where $F$ is the thermal factor of Eq.\ (\ref{self}).
Again using the angular integral
\begin{equation}
\int \frac{d\Omega _{q}}{(2\pi )^{3}}\mathrm{Im}G(\nu +\omega ,\mathbf{%
k+q)\propto }\frac{1}{v_{F}q},
\end{equation}%
we find
\begin{align}
\mathrm{Im}\Sigma (\omega )& \propto \lambda ^{2}\chi _{0}\int_{-\infty
}^{\infty }d\nu F(\nu ,\omega )\nu \int dq^{2}\delta (\nu -\omega
_{r}-aq^{2})  \notag \\
& \propto \lambda ^{2}\chi _{0}\int_{\omega _{r}}^{\infty }d\nu F(\nu
,\omega )\nu   \notag \\
& \propto \lambda ^{2}\chi _{0}T^{2}I_{1}(\frac{\omega _{r}}{T},\frac{\omega 
}{T})  \label{ImSigma_res}
\end{align}%
where $I_{1}(z,y)$\ has been defined in Eq.\ (\ref{ii}). The vertex correction $\lambda $
and the static spin susceptibility are both temperature dependent. Below, we
approximate $\chi $ in the relevant regime $0.3K<T<2K$ and $0<H<4T$\ by 
$\chi (T)\propto \ln (10/T)$ , which describes the data reasonably well.
The corresponding contribution to the magnetoresistivity will have soft threshold
behavior at $T_{2}^{\ast }(H)\approx \omega _{r}$. 

\section{ T*-line in the phase diagram of YRS}

We are now ready to collect our results on the location of the $T^*$ line in
the $T-H$ phase diagram of YRS. These are determined in two ways, with identical results: i) from the position of the
midpoint of the step feature in the magnetoresistance, Hall coefficient and
other quantities, and ii) from the solutions of the implicit equations $\pi T^{\ast }\approx u(T^{\ast },H)$ in
the low temperature regime (spin-flip scattering) $T\lesssim 0.3K$ and $T^{\ast }\approx \omega
_{r}(T^{\ast },H)$ at higher $T$( scattering from spin resonance).
The collected $T^*$ points from our calculations are shown
 in Fig.\ \ref{phase} together with the $T^*$ line published in
numerous papers by the Dresden group \cite{Silke,sci,fried10}. The agreement
is seen to be very good. 
As shown above, the step heights of the features associated with $T_{1}^{\ast}$ and $T_{2}^{\ast}$  vary with temperature as $T^{3/4}$ and $T^{2}$, respectively, which explains why  the $T_{1}^{\ast}$ feature is less important at higher $T$, and vice versa.

\begin{figure}[h]
\centering
\includegraphics[width=.45\textwidth, %viewport=30 80 600 10,%clip
]
{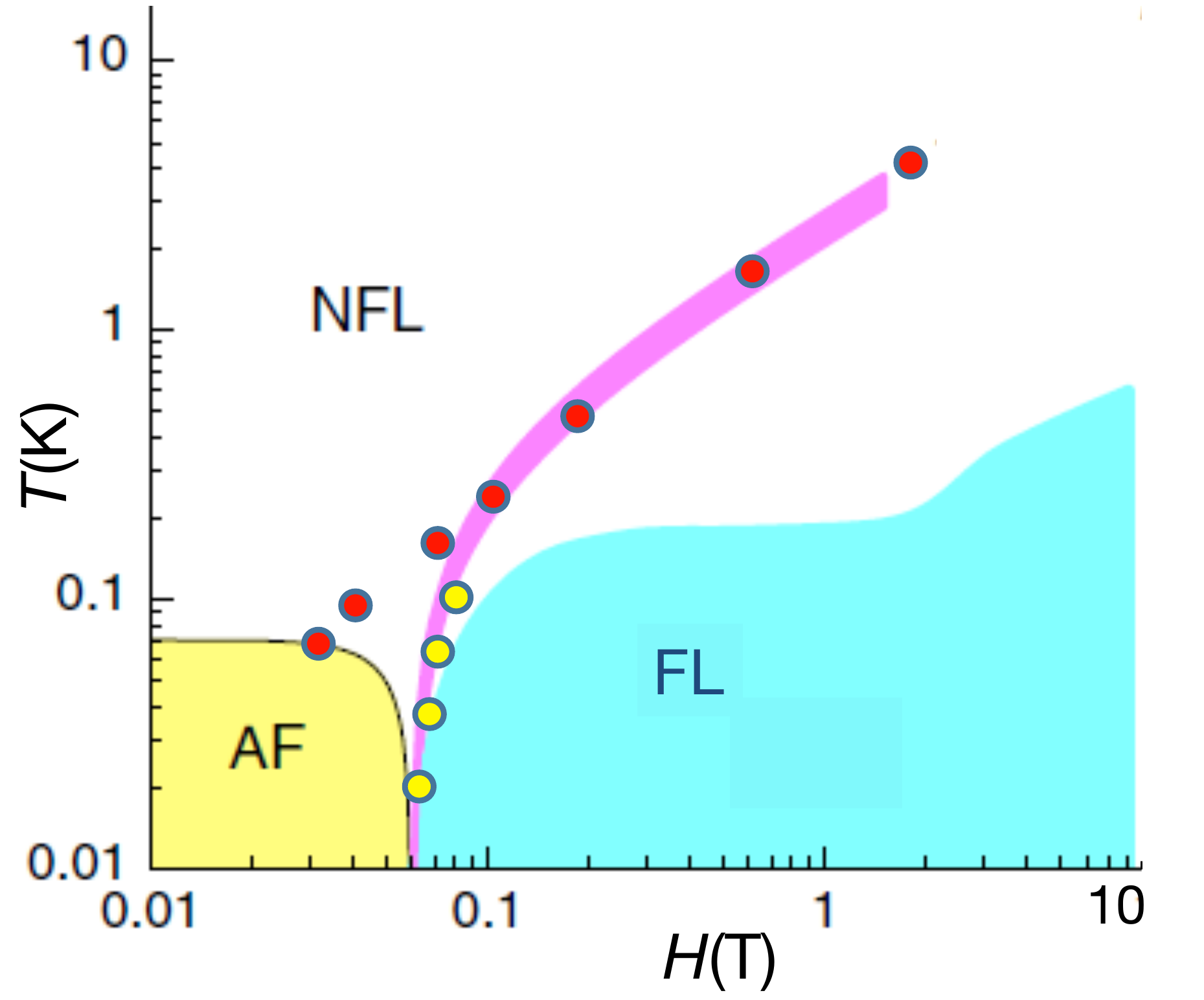}
%\vskip -1.5cm
\caption{Experimental phase diagram of YRS.\cite{gegenNJP} In the FL region (blue) the resistivity is $\propto T^2$. NFL denotes the non-Fermi liquid region where the resistivity varies with $T$  as $T^{\alpha}$ with $\alpha\leq 1$. The purple region is where several experimental probes exhibit a crossover behavior, called the $T^*$-line. The dots are the theoretical positions of the onset of spin-flip scattering -  from  quantum fluctuations (yellow) and at higher $(T,H)$, from  the spin resonance (red). The red dots at $H<0.1T$ are calculated using unpublished data \cite{scheff} }
\label{phase}
\end{figure}

\section{Conclusion}

We have addressed the crossover behavior in transport and thermodynamic quantities that occurs across the line $T^*(H)$ in the small $(T,H)$ region of the phase diagram of YRS. We propose  that the $T^*$ line marks the onset of spin-flip scattering processes. We show in detail how
these processes are switched on provided the temperature, and therefore the
thermal energy is sufficiently high to allow additional scattering processes
of at least two different types: (1) quasiparticle spin-flip
scattering off the quantum fluctuations associated with the QCP of YRS. This involves excitation over the Zeeman gap, which we show to nearly vanish
at the QCP; and (2) scattering off spin resonance bosons,
relevant at higher magnetic fields.
While the second contribution is noncritical and therefore
affects only the transport quantities, the first involves quantum
critical excitations and is therefore operative in both the transport and the
thermodynamic quantities. We have demonstrated that the observed
magnetoresistivity and the Hall coefficient may be quantitatively explained
by our model calculation. 

In our calculation of the magnetotransport properties, we have made extensive use, as input, of experimental data on specific heat, susceptibility and magnetization. This enables us to conclude that in the experimentally relevant temperature regime the rapid drop of the Zeeman splitting $h$ as the magnetic field is
lowered to below the critical field at fixed temperature is not so much
controlled by the decrease of  the quasiparticle weight $Z(H,T)$, but it is governed by the $H$ - dependence of the differential susceptibility and of the magnetization. Therefore the $T^{\ast }$-line is not necessarily tied to the critical
field (although at lower temperature it presumably is). This is to say that
if the QCP is shifted to higher or lower values of magnetic field by doping
the pure compound appropriately, this does not necessarily mean that the $%
T^{\ast }(H)$ as obtained above will follow the shift of the QCP. Rather,
it may stay approximately at the unshifted position. This may be easily
checked as soon as sufficient data on specific heat, magnetization and
susceptibility become available. The part of the $T^{\ast }$-line at higher
temperature, which according to our calculation is controlled by the scattering
off the spin resonance excitations will stay unchanged upon doping as long
as the resonance frequency is not affected by doping.

We emphasize that our description of the $T^*(H)$ line is simply based on incorporating spin-flip scattering from critical fluctuations and scattering from the spin resonance mode into the transport and thermodynamic responses. It does not depend on some consequence of a possible breakdown of Kondo screening near the critical magnetic field

\section{Acknowledgements}

We acknowledge useful discussions with H. v. L\"{o}hneysen, F. Steglich, J.
Thompson, A. Rosch, Q. Si, C. M. Varma, M. Vojta, and especially A. V.
Chubukov. Special thanks to Sven Friedemann, Philipp Gegenwart, Marc Scheffler, 
and J\"{o}rg Sichelschmidt for
sharing some of their experimental data with us. P.W. thanks the Department
of Physics at the University of Wisconsin--Madison for hospitality during a
stay as a visiting professor and acknowledges an ICAM senior scientist
fellowship. Part of this work was performed during the summers of 2012-14 at
the Aspen Center for Physics, which is supported by NSF Grant No.
PHY-1066293. P.W. acknowledges financial support by the Deutsche
Forschungsgemeinschaft through Grant No. SCHM 1031/4-1 and by the research
unit FOR960 `Quantum Phase Transitions'.


\begin{thebibliography}{99}
\bibitem{Doniach} S. Doniach, Physica 91B, 231 (1977).

\bibitem{LRVW} H. v. L\"{o}hneysen, A. Rosch, M. Vojta, and P. W\"{o}lfle,
Rev.\ Mod.\ Phys.\ \textbf{79}, 1015 (2007).

\bibitem{Silke} S. Paschen \textit{et al}, Nature \textbf{432} (2004).

\bibitem{sci} P. Gegenwart \textit{et al}, Science \textbf{315}, 969 (2007).

\bibitem{Si} Q. Si, S. Rabello, K. Ingersent, and J. L. Smith, Nature
(London) \textbf{413}, 804 (2001).

\bibitem{pc} P. Coleman, C. Pepin, Q. Si, and R. Ramazashvili, J. Phys.:
Condens. Matter \textbf{13}, R723 (2001).

\bibitem{SVS} T. Senthil, M. Vojta, and S. Sachdev, Phys. Rev. B \textbf{69}%
, 035111 (2004); M. Vojta, J. Low Temp. Phys.\textbf{161}, 203 (2010).

\bibitem{bs} S. Friedemann \textit{et al}, Phys.\ Rev.\ B \textbf{82},
035103 (2010).

%\bibitem{sven} S. Friedemann {\it et al}, Proc. Natl. Aca. Sci {\bf 107}, 14547 (2010).

\bibitem{zw} G. Zwicknagl, Phys. Rev. B

\bibitem{esr1} Elihu Abrahams and Peter W\"{o}lfle, Phys. Rev. B \textbf{78}%
, 104423 (2008).

\bibitem{esr2} Peter W\"{o}lfle and Elihu Abrahams, Phys. Rev. B \textbf{80}%
, 235112 (2009).

%\bibitem{WA1} Peter W\"{o}lfle and Elihu Abrahams, Phys. Rev. B \textbf{84}, 041101 (2011).

\bibitem{sichel} J. Sichelschmidt, V.A. Ivanshin, J. Ferstl, C. Geibel, and
F. Steglich Phys. Rev. Lett \textbf{91}, 156401 (2003).

\bibitem{schaufuss} U. Schaufuss, \textit{et al}, Phys. Rev. Lett. \textbf{%
102},\ 076405 (2009).

\bibitem{Broholm} C. Stock \textit{et al}, Phys. Rev. Lett. \textbf{109},
127201 (2012).

\bibitem{exp} J. Custers \textit{et al}, Nature \textbf{424}, 524 (2003).

\bibitem{asw14} E. Abrahams, J. Schmalian and P. W\"{o}lfle, Phys. Rev. B 
\textbf{90}, 045105 (2014).

\bibitem{wa11} Peter W\"{o}lfle and Elihu Abrahams, Phys. Rev. B \textbf{84}%
, 041101 (2011)

\bibitem{aw12} E. Abrahams and Peter W\"{o}lfle, Proc. Natl. Aca. Sci. 
\textbf{109}, 3238 (2012).

\bibitem{Hertz} J. A. Hertz, Phys. Rev. B \textbf{14}, 1165 (1976).

\bibitem{Millis} A. J. Millis, Phys. Rev. B \textbf{48}, 7183 (1993).

\bibitem{HvL} H. v. L\"{o}hneysen, \textit{et al}, J. Phys.: Condens.
Matter. \textbf{8}, 9689 (1996)

\bibitem{trov} O. Trovarelli \textit{et al}, Phys. Rev. Lett. \textbf{85},
626 (2000)

\bibitem{geg02} P. Gegenwart, et al., Phys. Rev. Lett. \textbf{89}, 056402
(2002).

\bibitem{geg06} P. Gegenwart, et al. , J. Phys. Soc. Jap. \textbf{75}, 155
(2006).

\bibitem{cust} J. Custers, et al., Phys Rev. Lett. \textbf{104}, 186402
(2010).

\bibitem{tok09} Y. Tokiwa, et al., Phys. Rev. Lett. \textbf{102}, 066401
(2009).

\bibitem{bran} M. Brando, et al., Phys. Stat. Sol. B \textbf{250}, 485
(2013).

\bibitem{zwick10} S. Friedemann, S. Wirth, N. Oeschler, C. Krellner, C.
Geibel, F. Steglich, S. MaQuilon, Z. Fisk, S. Paschen, and G. Zwicknagl,
Phys. Rev. B \textbf{82}, 035103 (2010).

\bibitem{pasch04} S. Paschen, T. L{\"u}hmann, S. Wirth,
P. Gegenwart, O. Trovarelli, C. Geibel, F. Steglich, P. Coleman, and Q. Si,
Nature \textbf{432}, 881 (2004).

\bibitem{fried10} S. Friedemann, N. Oeschler, S. Wirth, C. Krellner, C.
Geibel, F. Steglich, S. Paschen, S. Kirchner, and Q. Si, PNAS \textbf{107},
14147 (2010) and J.\ Phys: Condens.\ Matter {\bf 23}, 094216 (2011).

\bibitem{CeCuAu} Q. Si, J.-X. Zhu, and D. R. Grempel, J. Phys.: Condens.
Matter, \textbf{17}, R1025 (2005).

\bibitem{pep} I. Paul, C. P{\' e}pin and M.R. Norman, Phys. Rev. Lett. 
\textbf{98}, 026402 (2007); Phys. Rev. B \textbf{78}, 035109 (2008); K-S.
Kim and C. P{\' e}pin Phys. Rev. B \textbf{81}, 205108 (2010) and references
therein.

\bibitem{abch} Ar. Abanov and A.V. Chubukov, Phys. Rev. Lett. \textbf{84},
5608 (2000); Phys. Rev. Lett. \textbf{93}, 255702 (2004).

\bibitem{metsach} Max A. Metlitski and Subir Sachdev, Phys. Rev. B \textbf{82%
}, 075128 (2010). 
%\bibitem {MC}D. L. Maslov, and A. V. Chubukov, Phys. Rev. B \textbf{81},
%045110 (2010).

\bibitem{oesch} N. Oeschler \textit{et al}, Physica B \textbf{403}, 1254
(2008) and N. Oeschler, private communication.

%\bibitem{WA0} Peter W\"{o}lfle and Elihu Abrahams, Phys. Rev. B \textbf{80}, 235112 (2009).

\bibitem{geg} P. Gegenwart, Q. Si and F. Steglich, Nature Phys. \textbf{4},
186 (2008).

\bibitem{ishida} K. Ishida \textit{et al}, Phys.\ Rev.\ Lett.\ \textbf{89},
107202 (2002).

\bibitem{gegenNJP} Adapted from P. Gegenwart \textit{et al}, N. J. Phys. {\textbf 8}, 171 (2006).

\bibitem{scheff} M. Scheffler, and J. Sichelschmidt, private communication.

\end{thebibliography}
\end{document}